\documentclass[aps,a4paper,english,twocolumn ,prl, 10pt, superscriptaddress ]{revtex4}

\usepackage{amsmath,amsfonts,amssymb,dsfont,amsxtra,amsopn,amstext,amsbsy}

\usepackage[T1]{fontenc}
\usepackage[latin1]{inputenc}

\usepackage{color,cancel,wrapfig,lscape,array}
\usepackage{textcomp}
\usepackage{times}
\usepackage{ragged2e}
\usepackage{mathptmx}
\usepackage{latexsym}
\usepackage[all]{xy}
\usepackage{hhline}
\usepackage{enumerate}
\usepackage{bbold}
\usepackage{bbm}
\usepackage{bm}
\usepackage{dsfont}
\usepackage{units}
\usepackage{nicefrac}
\usepackage{graphicx}
\usepackage[dvips]{rotating}
\usepackage{color}
\definecolor{violet}{cmyk}{0.94,0.86, 0.04,0.08}
\usepackage[
colorlinks, 
linktocpage, 
bookmarksnumbered, 
bookmarksopen, 
pdfstartview=FitH, 
linkcolor=blue, 
citecolor=blue, 
urlcolor=magenta, 
filecolor=blue %
backref 
]{hyperref}

\newcommand{\nt}{\notag}

\newcommand{\dg}{\dagger}
\newcommand{\phdg}{{\phantom{\dagger}}}

\newcommand{\erz}[2]{\hat #1^\dagger_{#2}}
\newcommand{\ver}[2]{\hat #1^\phdg_{#2}}
\newcommand{\erw}[1]{\langle #1 \rangle}

\newcommand{\ii}{\mathbbmtt{i}}

\renewcommand{\vec}[1]{\bm{\mathbf{#1}}}

\newcommand{\qv}{{\vec{q}}}

\newcommand{\cc}{\text{c}}
\newcommand{\vv}{\text{v}}

\newcommand{\ee}{\mbox{\upshape e}}

\newcommand{\delt}{\partial_{t}}

\newcommand{\ham}{\hat H}
\newcommand{\hcpn}{\hat H_\text{el-pn}}
\newcommand{\hcpt}{\hat H_\text{el-pt}}

\newcommand{\cp}{\clearpage}

\begin{document}

\title{Stabilization of collapse and revival dynamics by a non-Markovian phonon bath}
\date{\today}
\author{Alexander Carmele}
\email [E-mail at: ]{alex@itp.physik.tu-berlin.de}
\author{Andreas Knorr}
\author{Frank Milde}
\affiliation{Institut f\"ur Theoretische Physik, Nichtlineare
Optik und Quantenelektronik,\\
Technische Universit\"at Berlin, Hardenbergstra{\ss}e 36,
EW 7-1 10623 Berlin, Germany}


\maketitle

\begin{bf}


Semiconductor quantum dots (QDs) have been demonstrated to be versatile candidates to study the fundamentals of light-matter interaction~\cite{Shields:NaturePhoton:07,deVasconcellos:NaturePhoton:10, Benson:Nature:11}. In contrast with atom optics, dissipative processes are induced by the inherent coupling to the environment and are typically perceived as a major obstacle towards stable performances in experiments and applications~\cite{Ladd:Nature:10}.

In this paper we show that this is not necessarily the case. In fact, the memory of the environment can enhance coherent quantum optical effects. 
In particular, we demonstrate that the non-Markovian coupling to an incoherent phonon bath has a stabilizing effect on the coherent QD cavity-quantum electrodynamics (cQED) by inhibiting irregular oscillations and boosting regular collapse and revival patterns. For low photon numbers we predict QD dynamics that deviate dramatically from the well-known atomic Jaynes-Cummings model. 
Our proposal opens the way to a systematic and deliberate design of photon quantum effects via specifically engineered solid-state environments. 
\end{bf}

Semiconductor QDs, coupled to a photonic cavity, exhibit signatures of strong coupling~\cite{Reithmaier:Nature:04} when the electron-photon (el-pt) interaction outrivals the combined dipole decay rate and cavity loss.
In contrast to fundamental atom-photon interfaces~\cite{Cirac:PhysRevLett:95}, a key factor to the understanding of observed phenomena in the semiconductor cQED regime is the interaction of electrons with the QD host material. 
The inherent many-body properties of a solid-state, in particular the electron-phonon (el-pn) interaction, lead to often undesired decoherence, where the superposition of otherwise distinct quantum states becomes lost and information about the system is carried away into the surrounding of the embedded QD~\cite{Carmele:PhysRevB:10}. In principle, this seems to put limits on the prospects of the wide field of quantum information processing and communication~\cite{Zoller:EurPhysJD:05}.
However, it has been shown that in photosynthetic systems, for example, the quantum-coherent excitation transfer in Fenna-Matthews-Olsen (FMO) antenna complexes is not only resistant to dephasing~\cite{Lee:Science:07}, but is actually supported by thermal fluctuations of the molecule~\cite{Mohseni:JChemPhys:08,Plenio:NewJPhys:08}. This sparked a lively debate over its implications for evolutionary biology, energy technologies and quantum information~\cite{Lloyd:NaturePhys:09}.  

It is also known that lattice vibrations in semiconductor nano structures can give rise to new effects not known in atomic quantum optics, nor in Markovian treatments of the semiconductor environment, e.g. phonon-mediated off-resonant cavity feeding~\cite{Hughes:PhysRevB:11,Hohenester:PhysRevB:09}, formation of phonon-assisted Mollow triplets~\cite{Kabuss:PhysRevB:11} and temperature-dependent vacuum Rabi splittings in cavity emission spectra~\cite{Milde:PhysRevB:08}.
Very recently impressive experiments were realized which could steer the transition from Markovian to non-Markovian dynamics and thus the flow of information between an open system and its environment~\cite{Liu:NaturePhys:11}.  

Expanding on these exciting developments we report on how non-Markovian coupling to an environment can be exploited to boost quantum optical features: 
The memory effects of an incoherent bath of harmonic oscillators induce astonishing pattern formations in the coherent collapse and revival (CnR) phenomenon of the QD electron density dynamics in a nanocavity. The CnR phenomenon results from the interference of different Rabi frequencies (Jaynes-Cummings ladder contributions). 
The presented effect  further challenges the widely adopted notion that the environmental coupling, responsible for decoherence, constitutes a substantial drawback that needs to be overcome. 
The very same coupling does not destroy, but rather facilitates and stabilizes the otherwise fragile quantum effect of CnR, in particular at the few-photon level, and hence lifts the restriction of the high photon numbers needed in atomic systems to initiate stable CnRs. 
  
To provide a basis for discussion of solid-state effects on an equal footing, we start by introducing an approach to a two-level system interacting with (i) photons and (ii) phonons, which encompasses the well-known Jaynes-Cummings model (JCM) and non-Markovian pure dephasing effects, respectively. 

\begin{figure}[tb!]
	\begin{center}
		\includegraphics[width=5cm]{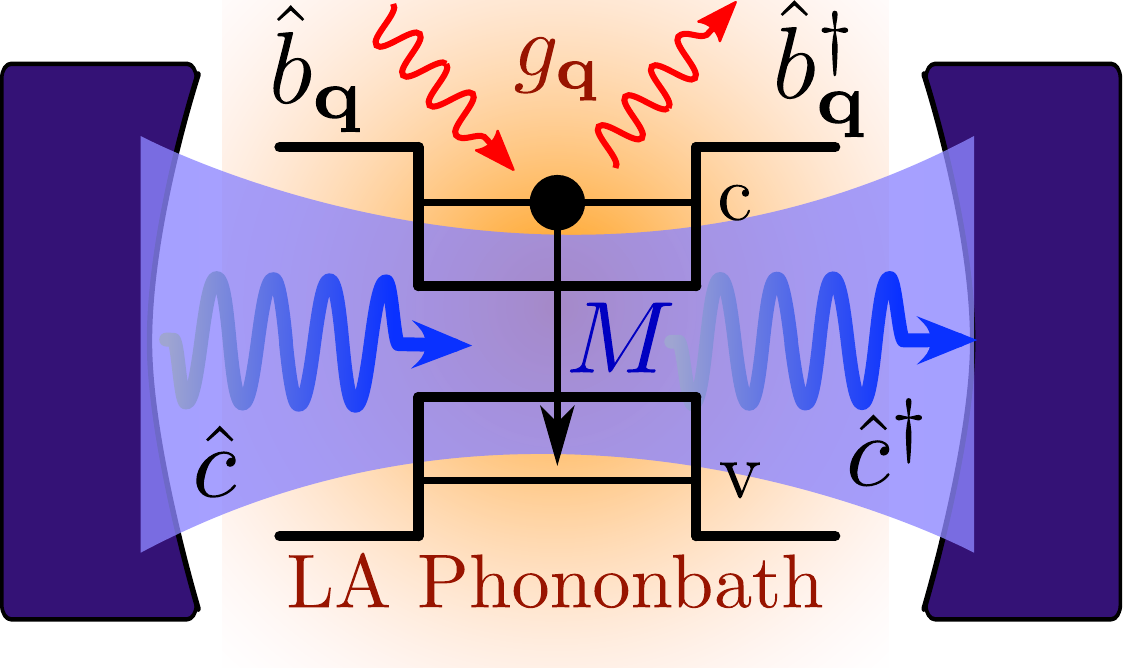}
		\caption[]{|~ ~{\footnotesize \textsf{\textbf{Investigated system.} A two-level QD with states v and c in a photonic cavity is coupled to (i) a single cavity mode (blue) via $M$ and (ii) a phonon bath (orange) via $g_\qv$.}}}
		\label{fig:scheme}
	\end{center}
\end{figure}

We consider an InAs/GaAs self-assembled QD with well separated quantum confined electronic states, i.e valence (v) and conduction (c) shell. This effective two-level system is embedded in a single-mode photonic cavity and can interact with both, the quantized lattice vibrations of the QD host matrix and the cavity photons, see Fig.~\ref{fig:scheme}.  
Particularly, we will focus on the interaction of electrons with longitudinal-acoustic (LA) phonons via deformation potential coupling, dominating at low temperatures~\cite{Vagov:PhysRevB:11,Forstner:PhysRevLett:03}.  The investigation will explore the temporal dynamics of the ground and excited QD states, since they are directly experimentally accessible in measurements of the photocurrent of a single-QD photodiode \cite{Zrenner:Nature:02} or extractable from the photon density of a cavity with output coupling \cite{Cao:ApplPhysLett:97}.

The dynamics of the relevant expectation values is derived via the Ehrenfest theorem 
\begin{equation}
	-\ii\hbar\delt\erw{\hat O(t)}=\langle [\hat H, \hat O]_- \rangle.
	\label{eq:ehrenfest}
\end{equation}
The Hamiltonian of our model includes the free energy of non-interacting electrons, cavity photons and bulk phonons:
\begin{equation}
	\hat H_0=\sum_i^\text{{v,c}}\hbar\omega_i \erz{a}{i}\ver{a}{i}+\hbar\omega_0 \erz{c}{}\ver{c}{} + \sum_\qv \hbar\omega_\qv \erz{b}{\qv}\ver{b}{\qv},
	\label{eq:h0}
\end{equation}
with electron creation (annihilation) operators $\ver{a}{i}(\erz{a}{i})$ in shell $i=$v,c. Usually, the respective quantized energies $\hbar\omega_i=\varepsilon_i$ are obtained within the  effective mass approximation. The frequency of the fundamental cavity mode is $\omega_0$ and $\ver{c}{} (\erz{c}{})$ create (annihilate) a corresponding photon. Phonons with wave vector $\qv$ are described by operators $\ver{b}{\qv}$ and $\erz{b}{\qv}$.  Their modes have a linear dispersion $\omega_\qv=c_\text{s} \cdot |\qv|$, where $c_\text{s}$ is the speed of sound. 
The el-pt interaction is treated within dipole and rotating wave approximation~\cite{Carmichael::99}:
\begin{equation}
	\hcpt=-\hbar M (\erz{a}{\vv}\ver{a}{\cc}\erz{c}{}+ \erz{a}{\cc}\ver{a}{\vv}\ver{c}{}),
	\label{eq:hamcpt}
\end{equation}
with an optical coupling strength of $M=\unit[100]{\mu eV}$. Finally, the band diagonal interaction of bulk phonons with the conduction shell QD electrons is given by 
\begin{equation}
	\hcpn=\sum_\qv g_\qv\erz{a}{\cc}\ver{a}{\cc}(\erz{b}{\qv}+\ver{b}{\qv}).
	\label{eq:hamelpn}
\end{equation}
Note, that the phonon bath is initially in equilibrium with the unexcited QD system. The coupling elements $g_\qv=g_\qv^\cc-g_\qv^\vv$ are given in the \emph{Methods} section.
In the same section, details are given on how to derive a general set of coupled differential equations, which allows on to determine the electron dynamics $\langle\erz{a}{\cc}\ver{a}{\cc} \rangle$. The relevant dynamical quantities needed are the photon-assisted ground and excited state density $G^m=\erw{\erz{a}{\vv}\ver{a}{\vv}\hat c^{\dg m}\hat c^m}$ $X^m=\erw{\erz{a}{\cc}\ver{a}{\cc}\hat c^{\dg m}\hat c^m}$, as well as the polarization $P^m=\erw{\erz{a}{\vv}\ver{a}{\cc}\hat c^{\dg m+1}\hat c^m}$. The number of involved photons is labeled by $m$.

We focus on a situation that starts with an inverted QD, i.e. $G^m(t_0)=0$ and $X^0(t_0)=1$, and a \emph{coherent} photon state with a mean photon number $N = 3.5$~\cite{Rempe:PhysRevLett:87}. Further details on parameters and initial conditions are found in the \emph{Methods} section as well. 

Before investigating the coupled electron, photon, and phonon dynamics, we benchmark our model and derive the exact solution of the coherent-state JCM~\cite{Narozhny:PhysRevA:81} by neglecting the el-pn interaction. For a cavity and QD on resonance Fig.~\ref{fig:density} (a) displays the corresponding Rabi oscillations in the occupation of the excited QD state $X^0$ showing their typical collapse at \unit[2]{ps} and subsequent revival within the first \unit[5]{ps}. However, due to the low initial photon number of $N=3.5$ in our model, the first CnR cycle is not completed and after \unit[10]{ps} the system drifts into irregular fluctuations with a standard deviation of $\sigma=0.026$ around the mean value of $0.5$, indicated by the shaded area. This behavior can be explained as follows:

\begin{figure}[tb!]
	\begin{center}
		\includegraphics[width=8cm]{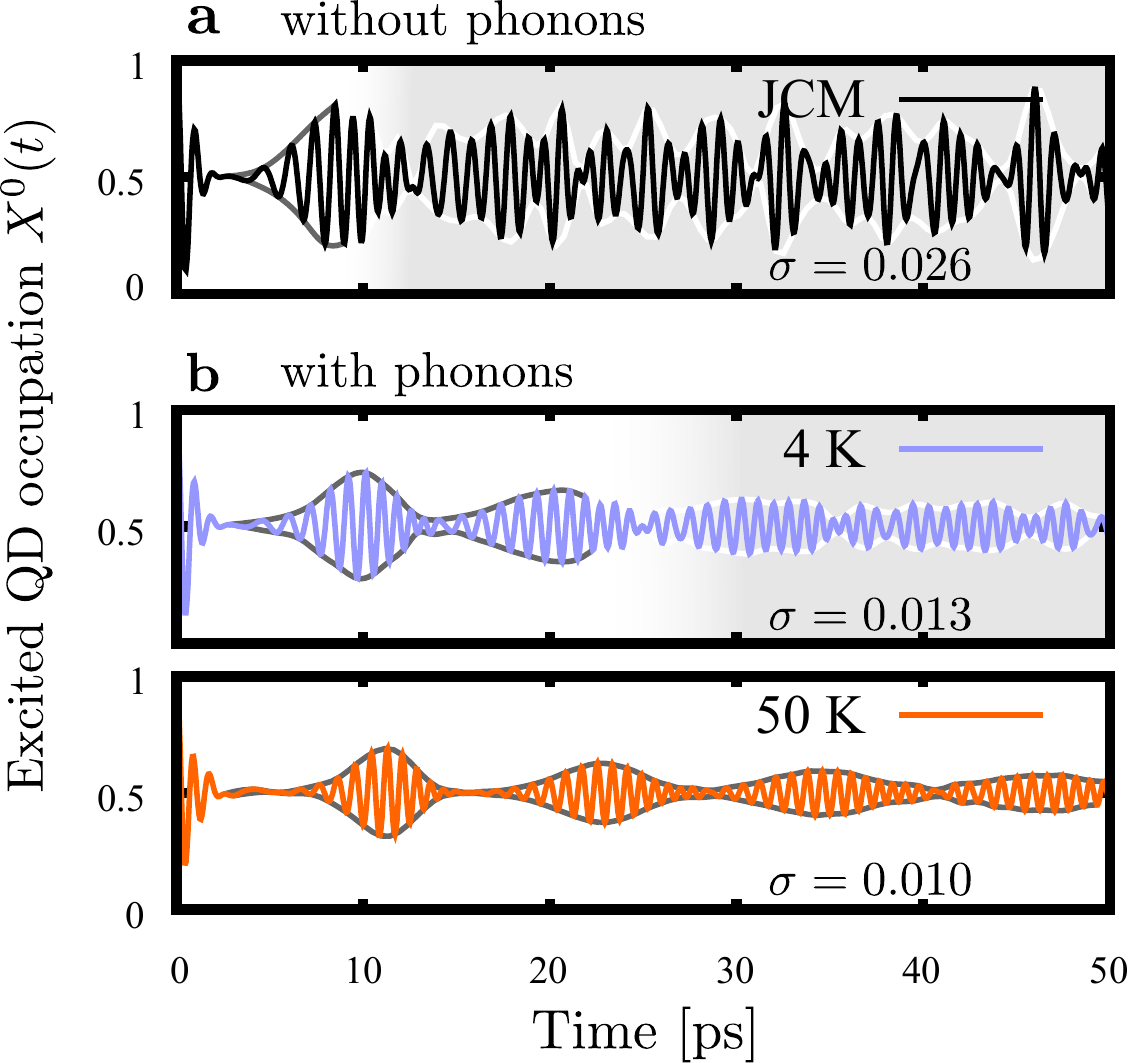}
		\caption[]{|~ ~{\footnotesize\textsf{\textbf{QD occupation for a mean photon number of $N=3.5$.} (a) Reproduced JCM solution without phonons shows only a single collapse with an incomplete revival. (b) Including phonons with a lattice temperature of \unit[4]{K} (blue) and \unit[50]{K} (orange)  many-cycle CnRs become possible. The shaded area indicates where oscillations are irregular.
		}}}
		\label{fig:density}
	\end{center}
\end{figure}

When the photon field is in a Fock number state with exactly $m$ photons and on resonance with the fundamental QD transition, the electronic density oscillates between the ground and excited state $S_m(t)=\cos^2(\Omega_m t)$ with the quantized Rabi frequencies (RFs) $\Omega_m= M \sqrt{m + 1}$. However, for an initially coherent field mode, expressed as a superposition of these number states, the typical Jaynes-Cummings solutions $S_m(t)$ for the initially occupied excited state $X^0$ are averaged over the probability distribution $P(m)$: $X^0(t) =\sum_m P(m) S_m(t)$. The Poissonian distribution $P(m) = \exp(-|\alpha|^2) |\alpha|^{2m}/m!$, which describes coherent photon states, exhibits a sharp peak at the mean photon number $|\alpha|^2 = N$ (in our simulation $N = 3.5$) and forms a quantum wave packet having certain reoccurrence times. Consequently, CnRs occur, cf. Fig.~\ref{fig:density} (a)~\cite{Narozhny:PhysRevA:81}. This behavior relates directly to the quantized nature of the cavity field $\left[c,c^\dg\right] \neq 0$: Typically, only discrete frequencies $\Omega_m$ around the mean photon number $N$ interfere, which are initially (at $t_0$) prepared in a definite state and therefore start a correlated dynamics. With increasing time the superimposed discrete RFs exhibit an increasing phase shift and destructive interference is inevitable. However, this collapse is followed by a revival of the excited state oscillation, when neighboring RFs differ by $2\pi$: $2 (\Omega_{m+1}-\Omega_m ) T_\text{R} \approx 2\pi$, with $T_\text{R}=2\pi M^{-1}\sqrt{N}$ as the revival time~\cite{Narozhny:PhysRevA:81}. The dynamics of this revival phenomenon strongly depends on the mean photon number: For high photon numbers, the impact of spontaneous emission processes is weak, i.e. the mixing of the superimposed RFs is negligible and the initially correlated state is recovered. As a result, many CnR events are visible (e.g. for $N=100$, not shown). If the mean photon number is small,  as in Fig.~\ref{fig:density} (a), the initial correlated state is not recovered and the excited state dynamics becomes highly irregular in a quasi-irreversible manner~\cite{Narozhny:PhysRevA:81}.

Now, we turn to the specifics of the light-matter coupling in a solid-state environment. 
When including phonons, the conduction shell electron density shows a strongly modified time evolution, see Fig.~\ref{fig:density} (b). Remarkably, the coupling to the phonon bath boosts coherent interference signatures: Depending on the temperature, CnRs survive for much longer times, e.g. for \unit[50]{K} up to \unit[80]{ps} (not shown here) before fading out. 
Finally, note that the fluctuation amplitude is greatly reduces by \unit[50]{\%} to a standard deviation of $\sigma=0.013$ for \unit[4]{K} and even lower to $\sigma=0.010$ at \unit[50]{K}. 

The counter-intuitive behavior of an incoherent bath of phonons supporting coherent features and suppressing seemingly random oscillations can be understood by investigating in detail the coherent polarization $P^m$ of the coupled el-pt-pn system, which drives the dynamics of $X^m$ and vice versa. 

\begin{figure}[tb!]
	\begin{center}
		\includegraphics[width=8cm]{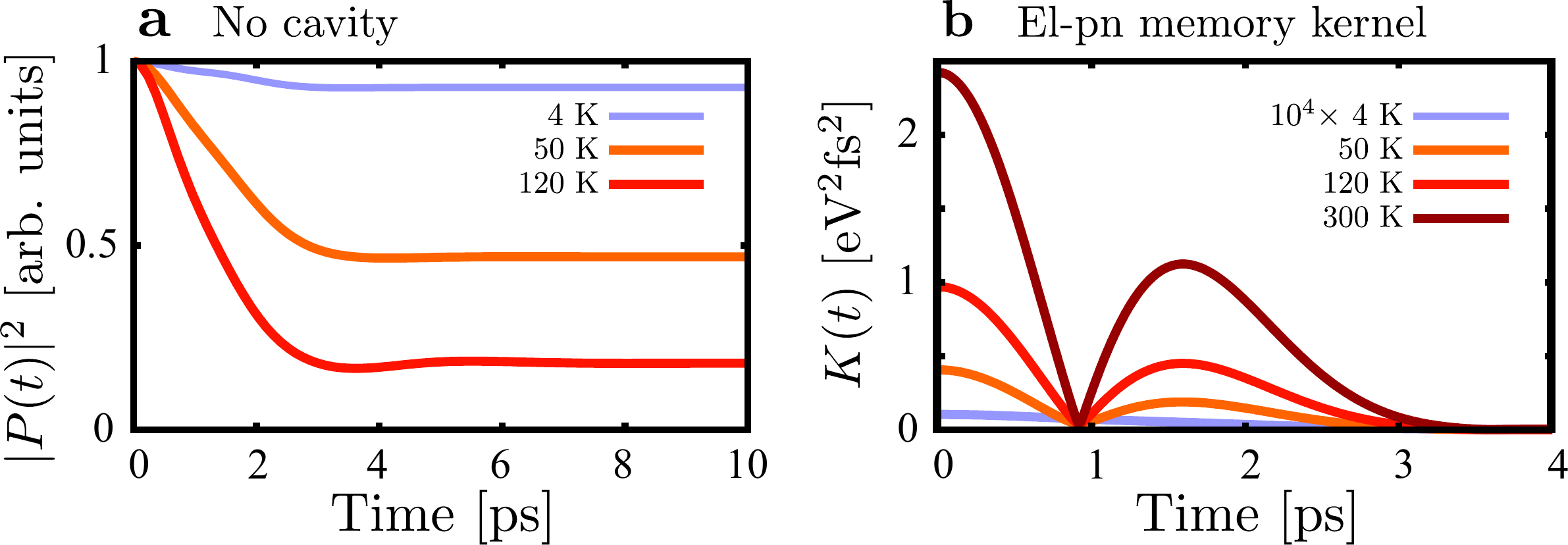}
		\caption[]{|~ ~{\footnotesize\textsf{\textbf{Electron-phonon coupling in QDs.} (a) Phonon-induced decay of the QD coherence \emph{in absence} of a cavity and hence without coupling to photons. (b) Absolute value of the corresponding el-pn memory kernel. Independent of temperature the kernel vanishes after \unit[4]{ps}. }}}
		\label{fig:memkernel}
	\end{center}
\end{figure}
As an instructive example, consider a simpler case of a QD that interacts solely with its host material and not with the quantum light field, first. Here, the band-diagonal coupling to the phonon bath will lead to so-called \emph{pure dephasing}, i.e. a destruction of the phase coherence of the bare initial two-level systems polarization $p=\langle \erz{a}{\vv}\ver{a}{\cc}\rangle$ without relaxation of the excited state occupation. 
Figure~\ref{fig:memkernel} shows how phonons induce an ultra-fast temperature-dependent decay of the polarization on a ps time scale. In contrast to a pure exponential decay, as obtained by introducing a simple $T_2$-time, the initial dephasing in Fig.~\ref{fig:memkernel} (a) reveals the non-Markovian nature of the system-bath coupling~\cite{Madsen:PhysRevLett:11} by a quadratic decay at $t=0$ and a clearly non-exponential behavior. Indeed, the memory kernel of the electron-phonon (el-pn) interaction~\cite{Vagov:PhysRevLett:07} persists for up to \unit[4]{ps}, independent of the bath temperature, see~Fig.~\ref{fig:memkernel} (b). 

\begin{figure}[tb!]
	\begin{center}
		\includegraphics[width=8cm]{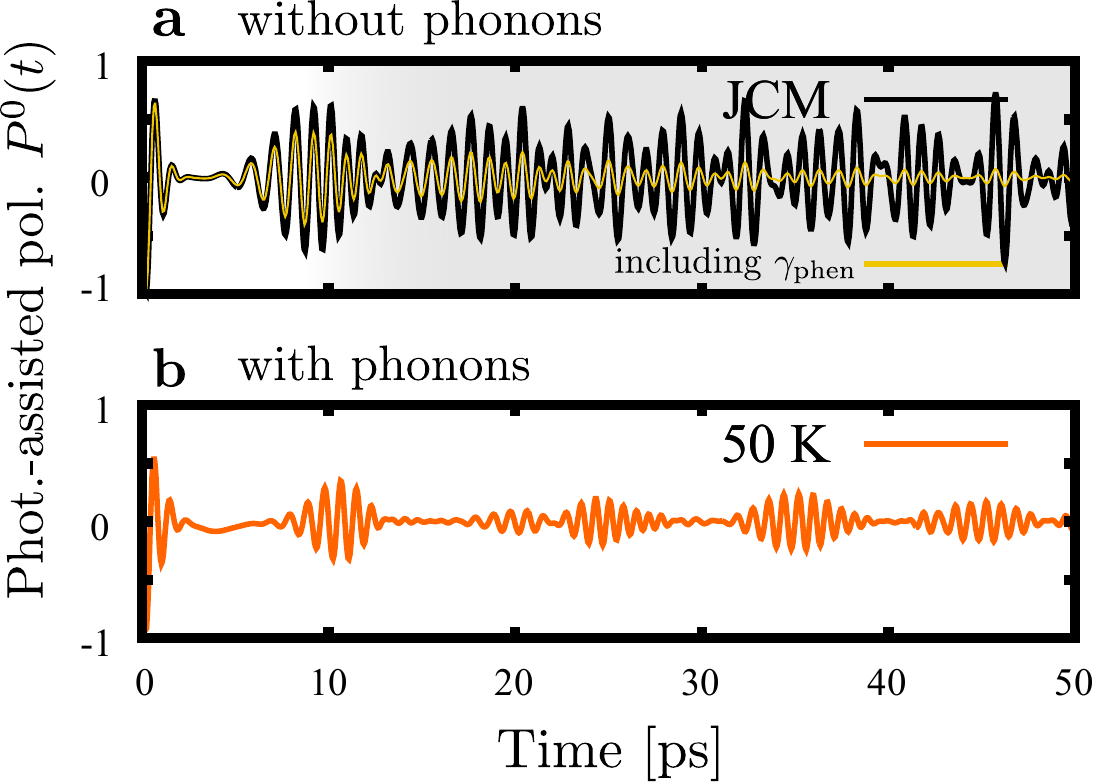}
		\caption[]{|~ ~{\footnotesize\textsf{\textbf{Coherent polarization dynamics.} The top panel (a) shows chaotic behavior of the JCM solution (black) without phonons. For comparison the dynamics including $\gamma_\text{phen}=\unit[0.1]{ps^{-1}}$ is shown as a yellow thin line on top of the black (undamped) JCM curve, not leading to many-cycle CnRs. The bottom panel (b) shows the phonon-induced dephasing of $P^0$ (orange) and resulting pattern formation.}}}
		\label{fig:pol}
	\end{center}
\end{figure}

Turning back to our cavity-QED model, the same decoherence happens to the photon-assisted coherences $P^m=\langle\erz{a}{\vv}\ver{a}{\cc}\hat c^{\dg m+1} \hat c^{m} \rangle$ via their coupling to the phononic system, see equation~\eqref{eq:p} in the \emph{Methods} part. These polarizations are the sources for the electronic CnRs. Exemplary, the temporal dynamics for $m=0$, i.e. $\langle \erz{a}{\vv}\ver{a}{\cc}\hat c^\dg\rangle$, is shown in Fig.~\ref{fig:pol}. When no phonons are considered (a) undamped and, at later times, chaotic oscillations (black line) occur after a first CnR event, continuously driving the photon-assisted electron densities $X^m$. 
In contrast, Fig.~\ref{fig:pol} (b) (orange line) shows how phonons affect the coherent dynamics by inducing a dephasing each time a revival emerges. However, due to the temporally finite interaction memory kernel, see Fig.~\ref{fig:memkernel} (b), $P^m$ can revive on a picosecond time scale.

In essence, the phonon bath neutralizes the phase shift between the neighboring RFs, which occurred during the el-pt dynamics, and effectively resets the RFs phase relation to recover the initial correlated state.  
Therefore, the initial CnR event is repeated and subsequent revivals can arise, only with a decreased amplitude, since the phonon bath introduces a loss channel in the system. 

We emphasize that this behavior is not seen when, instead of the full non-Markovian system-bath coupling, only a simple phenomenological dephasing constant $\gamma_\text{phen}$ is used, which resembles the Markovian contribution. The effect of a decay with $\gamma_\text{phen}=\unit[0.1]{ps^{-1}}$ is presented in Fig.~\ref{fig:pol} (a) by a thin yellow line on top of the undamped JCM solution (black line). Clearly, the oscillation amplitude gets damped. However, unlike including phonons, compare the orange line in Fig.~\ref{fig:pol} (b), no patterns in the chaotic fluctuations emerge, since a constant damping rate cannot re-phase the increasing shift between the RFs. Hence, only the non-Markovian nature of the el-pn coupling allows for a boost of the CnR phenomena.

Under what terms and conditions do we expect to observe this effect? Like for high mean photon numbers ($N>100$), a very strong el-pt coupling will suppress the phonon-induced gain of coherence, since the electron-bath interaction becomes negligible. Qualitatively, the stabilization of CnRs depends on the ratio between the light-coupling $M$ and the $\qv$-sum of the effective el-pn coupling $G:=\sqrt{\sum_\qv |g_\qv|^2}$.
If the effective el-pn coupling $G$ is larger than the RF $\Omega_N =M \sqrt{N + 1}$, revival times in the semiconductor are substantially elongated and longer observable, e.g. in GaAs $G/\Omega_N\le 50$. We predict a considerably large effect in the few-photon limit. 

Since the effective coupling $G$ depends on the material parameters, the temperature of the bulk, and geometry of the QD, semiconductor technology is capable of engineering CnRs with predefined, desired properties. Going even further, using the environment to prolong electronic coherences may give solid-state systems the critical technological edge to realize algorithms for coherent quantum information processing. Investigating quantum optics with QDs might even advance our understanding of energy transfer in more complex situations, e.g. photosynthesis.

To summarize, the influence of a bath of bulk LA phonons on the coherent dynamics of a QD two-level system in the cQED regime was explored. The well-known random oscillations in the electronic occupation, typical in the atomic JCM for low photon numbers, where altered significantly. Instead, due to LA phonons inducing a non-Markovian dephasing and thus resetting the coherence time, regular collapse and revivals arise and last for up to \unit[100]{ps} depending on temperature.  


\begin{footnotesize}
	{\normalsize\textbf{Methods}}

\textbf{Dynamics. }
	The general set of coupled equations that determine the electron dynamics are obtained via a method of induction~\cite{Kabuss:PhysStatusSolidiB:11}, when using the complete Hamiltonian $\ham = \ham_0+\hcpt+\hcpn$ and equation~\eqref{eq:ehrenfest}:

\begin{align}
	\delt  G^m =& \ii M [\hat P^{m+1}-(  P^{m+1})^* + m  P^{m-1} - m (  P^{m-1})^*],\label{eq:g}\\
	\delt  X^m =& -\ii M  P^{m+1} + \ii M (  P^{m+1})^*\label{eq:e},\\
	\delt  P^m =& -\ii (\Delta -\ii \gamma_\text{phen})  P^m -\ii M(m   X^{m-1} +   X^{m+1} -   G^{m+1})\label{eq:p}\\
	& -\ii\sum_\qv g^*_\qv   P^m_+(\qv) + g_\qv   P_-^m(\qv).\nt
\end{align}
The (photon-assisted) ground state density $G^m=\erw{\erz{a}{\vv}\ver{a}{\vv}\hat c^{\dg m}\hat c^m}$ is solely driven by the photon-assisted polarization $  P^m=\erw{\erz{a}{\vv}\ver{a}{\cc}\hat c^{\dg m+1}\hat c^m}$ via the el-pt coupling, as is the excited state density $X^m=\erw{\erz{a}{\cc}\ver{a}{\cc}\hat c^{\dg m}\hat c^m}$. The polarization experiences a free rotation due to a possible detuning $\Delta=\omega_\text{g}-\omega_0$ of the QD gap energy $\omega_\text{g}=\omega_\cc-\omega_\vv$ and cavity resonance. We can furthermore account for additional broadenings, observed in experiments, by introducing $\gamma_\text{phen}$~\cite{Laucht:PhysRevLett:09}. However, a cavity loss $\kappa$ is \emph{not} included here, as it merely reduces the amplitude of all involved quantities equally, which has no relevance to our principle findings.  

Besides the spontaneous emission of photons $(\propto m  X^{m-1})$, crucial for Rabi oscillations,  equation~\eqref{eq:p} includes a coupling to phonon-assisted polarizations $  P^m_+(\qv)=\erw{\erz{a}{\vv}\ver{a}{\cc}  c^{\dg m+1}  c^m \erz{b}{\qv}}$ and $  P^m_-(\qv)=\erw{\erz{a}{\vv}\ver{a}{\cc} \hat c^{\dg m+1}\hat c^m \ver{b}{\qv}}$, too. The dynamics of $P_\pm$ is solved within a Born approximation, where electronic and phononic variables factorize. This yields a closed set in terms of the el-pn coupling
\begin{align}
	\delt  P^m_+ (\qv)&= \ii(\Delta + \omega_\qv) P^m_\pm(\qv)-\ii n_\qv P^m(\qv) ,\\
	\delt  P^m_- (\qv)&= \ii(\Delta - \omega_\qv) P^m_-(\qv)-\ii(n_\qv+1) P^m(\qv),
	\label{eq:ppm}
\end{align}
where the mean phonon number $n_\qv=\erw{\erz{b}{\qv}\ver{b}{\qv}}$ occurs. 
For temperatures well below \unit[100]{K}, as considered here, a 
second-order Born approximation is well validated to solve the occurring hierarchy problem in the el-pn coupling, as can be seen by comparison to exactly solvable models~\cite{Forstner:PhysRevLett:03} and experiments. To explore the high-temperature regime, terms beyond second-order must be included or other approaches employed~\cite{Vagov:PhysRevB:11}.

The phonons of the homogeneous semiconductor material are modeled as a reservoir of harmonic oscillators, which introduces a temperature dependence to the dynamics via the thermal Bose-Einstein distribution function  $n_\qv(T)=(\exp[\hbar\omega_\qv/\{k_\text{B}T\}]+1)^{-1}$. The coupling to the bath will affect the coherent properties of the photon-assisted polarizations $P^m$ over time and lead to a non-Markovian pure dephasing, which cannot be accounted for by introducing the simple, constant dephasing time $T_2=\gamma_\text{phen}^{-1}$ alone. 

The complete set of equations equation~(\ref{eq:g}-\ref{eq:ppm}) is, however, not closed in terms of the photon number ($m$). Therefore, depending on the coupling strength $M$, a sufficient high order in $m$ is taken into account to reach convergence in the calculations. Thus the obtained solution is numerically exact and renders, in the absence of el-pn coupling, the JCM~\cite{Carmele:PhysRevLett:10}.

To have a fair comparison to the JCM, we  accounted for the polaron shift in the QD resonance $\Delta_\text{pol}=\sum_\qv |g_\qv|^2/\omega_\qv$ by accordingly detuning the cavity, so that both, QD and cavity, are on resonance again. If $\Delta_\text{pol}$ would not have been compensated, a higher RF and smaller amplitude would have resulted. We carefully checked our model for detuning effects: Including a strong phenomenological dephasing $\gamma_\text{phen}$ allows for a Fourier transformation. A spectral analysis of $P^0(\omega)$ showed that the involved RFs in both, the phonon-free JCM and the phonon cases, are essentially in the same spectral range.

\textbf{Initial conditions. }
The initial state of the photon expectation values is derived in dependence of the (given photon statistics and) mean photon number $N$. The cavity is prepared in a coherent photon state with a mean photon number $N = 3.5$ and thus $\erw{\hat c^{\dg m}\hat c^m}=\sum_n N^n (\ee^N n!)^{-1}\langle n| \hat c^{\dg m}\hat c^m| n \rangle = N^m$. Initially the QD is inverted, i.e. $G^m(t_0)=0.0$ and $X^0(t_0)=1.0$. The complete density operator $\rho$ considers the electron- $\rho^\text{el}$ and photon system $\rho^\text{pt}$, and reservoirs $\rho^\text{r}$ (LA phonons and dissipative photon modes). Initially at time $t_0$, $\rho$ factorizes $\rho(t_0) = \rho^\text{el}(t_0)\otimes\rho^\text{pt}(t_0)\otimes\rho^\text{r}(t_0)$. Consequently, the photon-assisted excited state factorizes $X^m(t_0)=X^0(t_0) N^m$. No coherence is present $P^m(t_0)=P^m_\pm(t_0)=0$. The initial state of the phononic bath is determined by the temperature of the system.

\textbf{Numerical parameters. }
Phonon coupling element $|g_{\vec q}|=|g^{\vec q}_{\vv}-g^{\vec q}_{\cc}|$, where $g^{\vec q}_{i}=\sqrt{\frac{\hbar q}{2 \rho c V}}D_{i}e^{-\frac{q^{2}\hbar}{4 m_{i}\omega_{i}}}$. Used parameters: photon order $m=27$, electron-photon coupling $\hbar M=\unit[0.0015]{fs^{-1}}$, sound velocity of GaAs $c_\text{s}=\unit[5110]{m/s}$, deformation potentials $D_{\vv}=\unit[-4.8]{eV}, D_{\cc}=\unit[-14.6]{eV}$, effective masses: $ m_{\cc} = 0.067$, $m_{\vv} = 0.45$, confinement energies $\hbar\omega_{\cc}=\unit[0.030]{eV}$, $\hbar\omega_\vv=\unit[0.024]{eV}$ and mass density of GaAs $\rho=\unit[5370]{kg/m^3}$.

\end{footnotesize}

\noindent
\textbf{References}


\noindent
\textbf{Acknowledgments}

\noindent
\begin{footnotesize}
We would like to thank Simon M\"uller for discussions. This work was financially supported by the Deutsche Forschungsgemeinschaft within the Sonderforschungsbereich 787 ``Nanophotonik''.
\end{footnotesize}

\noindent
\textbf{Author contributions}

\noindent
\begin{footnotesize}
A.C. worked on the inductive method and carried out the simulations. F.M. worked out the el-pn effects and wrote the paper with input from A.K.. A.K. guided the work. 
\end{footnotesize}

\noindent
\textbf{Additional information}

\noindent
\begin{footnotesize}
The authors declare no competing financial interests.
\end{footnotesize}

\end{document}